# Binding energies: new values and impact on the efficiency of chemical desorption

V. Wakelam[1], J.-C. Loison[2], R. Mereau[2], M. Ruaud[1]

1 Laboratoire d'astrophysique de Bordeaux, Univ. Bordeaux, CNRS, B18N, allée Geoffroy Saint-Hilaire, 33615 Pessac, France
2 Institut des Sciences Moléculaires de Bordeaux (ISM), CNRS, Univ. Bordeaux, 351 cours de la Libération, 33400, Talence, France

## ABSTRACT

Recent laboratory measurements have confirmed that chemical desorption (desorption of products due to exothermic surface reactions) can be an efficient process. The impact of including this process into gas-grain chemical models entirely depends on the formalism used and the associated parameters. Among these parameters, binding energies are probably the most uncertain ones for the moment. We propose a new model to compute binding energy of species to water ice surfaces. We have also compared the model results using either the new chemical desorption model proposed by Minissale et al. (2016) or the one of Garrod et al. (2007). The new binding energies have a strong impact on the formation of complex organic molecules. In addition, the new chemical desorption model from Minissale produces a much smaller desorption of these species and also of methanol. Combining the two effects, the abundances of CH3OH and COMs observed in cold cores cannot be reproduced by astrochemical models anymore.

## 1 INTRODUCTION

It is now well established that the surface of interstellar dust plays a crucial role for the formation of many chemical species starting with $H_2$. With the revision of some rate coefficients for gas-phase reactions (Luca et al. 2002; Geppert et al. 2006), it appeared impossible to form methanol ($CH_3OH$) in the gas-phase only whereas it was quite easy to form large quantities of this molecule through Langmuir-Hinshelwood reactions on the dust surfaces even at low temperature (Garrod et al. 2006). In order to reproduce the ~ $10^{-9}$ abundance observed in cold cores (Pratap et al. 1997), an efficient non-thermal desorption mechanism is needed to bring it back to the gas-phase. The observation of complex organic molecules in cold and dense environments of the interstellar medium has increased the interest of the community onto such processes (Bacmann et al. 2012; Cernicharo et al. 2012; Vastel et al. 2014).

Several non-thermal desorption mechanisms have been considered. The partial or entire heating of grains due to cosmic-ray particle collisions has been investigating by many authors (Leger et al. 1985; Hasegawa & Herbst 1993; Willacy & Millar 1998; Shen et al. 2004). Photo-evaporation due to direct UV photons or photons induced by cosmic-rays has also been proposed and intensively studied experimentally. For the recent measurements, it seems that the photo-evaporation is an indirect process for some molecules and that the efficiency depends on the specie, the nature of the surface, and the wavelength of the impacting UV photons (Muñoz Caro et al. 2010; Fayolle et al. 2011; Bertin et al. 2012, 2013; Fayolle et al. 2013). Another non-thermal desorption that has been proposed is associated with the energy released by the formation of $H_2$ on the surface (Roberts et al. 2007). The formation of $H_2$ at the surface of the grain would be so exothermic that it would result in the partial desorption of the mantles. The fraction of the mantle evaporating remains an unconstrained parameter in the models. Generalizing this idea, the process of chemical desorption has been proposed by Garrod et al. (2007). For any exothermic reaction occurring at the surface of the grains, the released energy could be transferred to the products and induce their evaporation. Although the mechanism has been included in astrochemical models (for instance Garrod et al. 2007; Vasyunin & Herbst 2013; Wakelam et al. 2014), its efficiency has been a free parameter. Minissale et al. (2016) recently conducted experiments of this process and found that the efficiency decreases with the ice coverage of the surface. Although they could do some measurements only on a small number of systems, they also proposed a new formalism to include this mechanism into gas-grain astrochemical models. Both formalisms from Garrod et al. (2007) and Minissale et al. (2016) depend on the binding energy of the species to the surface. In this paper, we propose a new model to compute the binding energies for species on water ice surfaces and investigate the new chemical desorption model proposed by Minissale et al. (2016).

The paper is organized as follows. Section 2 contains a general description of our gas-grain chemical model, the methods to compute the branching ratios of chemical desorption from Garrod et al. (2007) and Minissale et al. (2016), and the new model to compute binding energies. The list of new binding energies is available in the appendix. In Section 3, we compare our model results for cold cores conditions obtained with the two chemical desorption models, with and without the updated binding energies. We then conclude in the next section.

## 2 CHEMICAL MODELING

### 2.1 Model description

The chemical composition of the gas and the dust icy mantles is computed with the gas-grain code Nautilus described in Ruaud et al. (2016). The reactions considered for the gas-phase chemistry are listed in the public chemical network kida.uva.2014 (Wakelam et al. 2015). The surface network is based on the one of Garrod & Herbst (2006). The chemical model includes physisorption of gas-phase species on grain surfaces, diffusion of species at the surface of the grains resulting in chemical reactions and several desorption mechanisms. The species on the surface can be desorbed due to the temperature (thermal desorption), cosmic-ray heating (cosmic-ray induced desorption, Hasegawa & Herbst 1993), UV photon impact (photodesorption) and chemical desorption (see section 2.2). The surface chemistry is solved using the rate equation approximation and assuming a different chemical behaviour between the surface of the mantle and the bulk (three-phase model), the surface and the bulk being both chemically active. All details and equations (except for the chemical desorption which is the subject of this paper) are given in Ruaud et al. (2016).

### 2.2 Computation of chemical desorption branching ratios

The chemical desorption mechanism is based on the idea that the energy released by exothermic reactions at the surface of the grains is partly transferred to the produced species. This energy is then distributed over the degrees of freedom of the molecule. The part of the energy that goes to the direction perpendicular to the surface will allow for the molecule to bounce on the surface with a probability to result in the desorption of the species. The efficiency of this process depends on the amount of energy that stays in the product and is not lost in the grain. This last parameter depends on the surface and is quite uncertain. In all cases, we assume that if the reaction results in more than one product, the chemical desorption is not efficient since the energy would then be distributed in the two products (see Garrod et al. 2007).

Two formalisms have been proposed in the literature to include this process in chemical models. Garrod et al. (2007) have used the theory of Rice-Ramsperger-Kessel (Rice & Ramsperger 1927, Kassel 1928), in which the probability of desorption is expressed as

$$P = \left(1 - \frac{E_D}{E_{reac}}\right)^{s-1}$$

with $E_D$ the binding energy of the product, $E_{reac}$ the energy released by the reaction (enthalpy of reaction), and $s$ the number of vibrational modes in the molecule/surface-bound system. This last parameter is equal to 2 is the product of the reaction is a diatomic species while it is $3n - 5$ (with $n$ the number of atoms in the specie) for other species. From $P$, we compute the fraction of products that would desorb at the end of the reaction by:

$$f = \frac{aP}{1 + aP}$$

with $a = \nu / \nu_S$ , and $\nu$ is the surface-molecule bond-frequency and $\nu_S$ the frequency at which the energy is lost to the grain surface. The value of $a$ is unknown and most studies consider values between 0.01 and 0.1, identical for all species, which leads to values of $f$ approximately equal to that of $a$, i.e. between 1 and 10%. We will call this formalism RRK in the rest of the paper.

More recently, Minissale et al. (2016) have proposed a new formalism, in which the fraction of evaporation depends on the mass of the product based on experimental results of a few surface reactions such as hydrogenation of O and CO, O + O and N + N reactions. In their formalism, the fraction of products evaporated is

$$f = e^{-\frac{E_D}{\epsilon E_{reac}/N}}$$

with N the number of degree of freedom's of the produced molecule ($N = 3n$) and $\epsilon = \frac{(M-m)^2}{(M+m)^2}$ is the fraction of the energy kept by the product with a mass $m$. $M$ is the effective mass of the surface, which depends on the nature of the surface and is not well constrained. Minissale et al. (2016) have shown that chemical desorption on water ices was much less efficient than on bare silicate or graphite grains. In fact, for most studied systems, the efficiency was below the detection level of the experiment. We have thus used the recommendation by these authors: 1) we have computed $f$ assuming the surface effective mass for bare grains of 120 amu and divided the obtained values by 10 and, 2) for the three systems where the chemical desorption could be measured, we used the measured values ($f_{O+H}$ = 30%, $f_{OH+H}$ = 25% and $f_{N+N}$ = 50%). We will call this formalism MDCH (for the names of the authors) in the rest of the paper. Except for the measured systems, this formalism produces evaporation fractions smaller than 10% and negligible fractions for most systems as $f$ decreases very rapidly with the binding energy of the product. Using the equations of Garrod et al. (2007), $f$ still decreases with $E_D$ but much less rapidly so that all surface reactions lead to the partial evaporation of the products. The percentage of product evaporation, for a selection of surface reactions, is given in Table 1 using the two formalisms. For both formalisms, the chemical desorption is only active for the surface layer.

## 2.3 Update of binding energies

Whatever the formalism chosen for the chemical desorption, its efficiency depends on the binding energy of the product. The binding energies on Amorphous Solid Water (ASW) are not well known, particularly for radicals. An estimation of these binding energies is usually derived from temperature-programmed desorption (TPD) experiments. This technique has been widely used to determine the binding energies of stable molecules and has been reviewed by Burke & Brown (2010) and Hama & Watanabe (2013). We partly complete these reviews in Table 2 including recent measurements for some atoms (H, O and N) for which the uncertainties are large. For species without measurements, there are various theoretical ways to calculate the binding energy with Ice, either through periodic representation of Water Ice or considering the Ice as a Cluster. However, these calculations are relatively complex and astrochemical models use rather additive law considering, for example, that $E_D(O_3) = E_D(O_2) + E_D(O)$, which is a very rough approximation (see for instance Cuppen & Herbst 2007). To estimate the unknown binding energies (for most of the radicals for example), we have developed a model founded on the stabilization energy of the complex between the various species and one water molecule. Then, we assume that the binding energy of the species with ASW is proportional to the energy of interaction between this species and one water molecule. To determine the proportionality coefficients, we fit the dependency of the experimental binding energies versus the calculated energies of the complexes for 16 stable molecules as shown in Fig. 1. The energies of the complexes were calculated at DFT level using the M06-2X functional (Zhao Y. & Truhlar D., 2008) associated to aug-cc-pVTZ level basis using Gaussian09, optimizing every degree of freedom but without inclusion of the ZPE nor BSSE correction. We performed various other ways to calculate the energy of the complexes including ZPE and BSSE corrections, as well as without optimizing the water geometry (to simulate the fact the geometry of water molecules in Ice is constrained by the structure of the Ice). We find that the different ways of calculating the complex energies had only a small influence on the quality of the fit (only changing the value of the parameters of the fit), the fit being slightly better without inclusion of the ZPE energy on the dimer. The high correlation factor of the fit (R= 0.97) strongly suggests that this is a good way to quickly estimate the binding energy of any species. It should be noted that we also performed a model using MP2/aug-cc-pVTZ calculations leading to similar results except for N, $CH_2$ and $CH_3$, the MP2 calculations leading to smaller binding energies.

The modelled (DFT and MP2 ones) binding energies are given in Table 2 in the appendix (columns 3 and 4). Except in very few cases (C, CH, $C_2$, Si, SiH species), the interaction between radicals and the water molecule does not involve strong (partly) covalent bonding. Instead, it is only a long range interaction (dispersion and dipole-dipole interactions) and the binding energy of radicals should be well described by our model, as for molecules. It should be noted that in various cases, the interaction is not isotropic, as it will be the case if only dispersion was involved in the interaction, but instead one geometry is favoured. In some cases, OH, $NH_2$, HNO, $SO_2$, and $HO_2$, there are various stable geometries for the complexes with very different interaction energies (see Table 2). In that case we use the average of the different values or we favour the geometry where the species interacting with H atom rather with O atom as ASW is supposed to have high concentration of dangling H atom, which are more accessible than oxygen atoms. The fact that there are in some cases multiple stable geometries for the complexes is likely correlated to the fact that there are various adsorption sites on Ice, and then the binding energy is not always well represented by a single value. $H_2$ is not included in the fit as in that case the energy of the complex between $H_2$ and one water molecule is highly depending of the method of calculation.

## 3 RESULTS

To show the impact of the chemical desorption on the chemical composition of cold cores, we have run Nautilus for a constant gas and dust temperature of 10 K, a density of $2 \times 10^4$ cm$^{-3}$, a visual extinction of 15, and a cosmic-ray ionization rate of $1.3 \times 10^{-17}$ s$^{-1}$. All species are initially atomic except for hydrogen, which is entirely molecular. Initial abundances are the same as in Ruaud et al. (2016).

### 3.1 Comparison of chemical desorption formalisms

To first investigate the effect of the formalism itself, we have run the model with the same binding energies (before updates) but using the two formalisms presented in Section 2.2. For the RRK, we have used an $a$ parameter of 0.01. The impact on the modelled abundances is not very strong except for large molecules strongly bonded on ice such as methanol and $CH_3NH_2$. Indeed, MDCH model strongly decreases the chemical desorption efficiency of large molecules because the efficiency is proportional to $e^{-n}$ with $n$ the number of atoms of the produced species. Fig. 2 shows the gas phase abundances of $CH_3OH$, $CH_3NH_2$, and $CO_2$ computed as a function of time using the two formalisms. The gas phase methanol is produced by the reactions

(1) $CH_2OH_{ice} + H_{ice} \rightarrow CH_3OH$ and

(2) $CH_3O_{ice} + H_{ice} \rightarrow CH_3OH$.

The fraction of the products that evaporate is approximately $2 \times 10^{-3}$ for each reaction with RRK while it is ten times smaller in the case of MDCH. In the case of $CH_3NH_2$, the main reaction of formation at $10^5$ yr is $CH_2NH_{2ice} + H_{ice}$, with a $f$ of $10^{-3}$ with RRK and 50 times smaller with MDCH. The gas phase abundance of $CH_3NH_2$ obtained with RRK reaches a detectable limit in cold cores. Observations of this molecule (resulting in a detection or not) in these regions could help constrain the chemical desorption during the formation of this species on the surface. With the MDCH model and the experimental methanol-Ice binding energy (from Collings et al. 2004), we cannot reproduce the methanol observed in dense molecular clouds contrary to the RRK model. This means either that our model underestimates the methanol formation on Ice, that the MDCH model is not valid for methanol desorption, or that we miss another non-thermal desorption process. Species (with an abundance larger than $10^{-12}$ as compared to the total proton density) affected by more than a factor of 3 at $10^6$ yr are: HOOH, $C_2H_6$, $CH_3CHO$, $CH_3CCH$, $CH_2CHC_2H$, $CH_3OCH_3$, $CH_3CO$, $CH_3O$ and $O_3$. All these species are mainly formed on the surfaces and chemically desorbed in the gas phase, and the fraction of products evaporated is smaller with MDCH than with RRK. For light species such as OH and NH and despite the larger *f* given by MDCH, the abundances are not significantly changed. The $CO_2$ abundance in the gas-phase shown in Fig. 2 is not much affected since the fraction of evaporation is about the same using both formalisms (0.9 for RRK and 1.8 for MDCH).

### 3.2 Model results with the updated binding energies

Using the formalism of MDCH, we have updated the code with the binding energies listed in Table 2. The new binding energies have an impact on the desorption of the species but also their diffusion since the diffusion energy is considered to be a fraction of the binding energy. The largest impact of these updates seems to come from the new experimental binding energy of atomic oxygen. In previous models, we assumed $E_D(O)$ to be 800 K based on Tielens & Hagen (1982) whereas new estimates seem to indicate that atomic oxygen is strongly bound to the surface (He et al 2005, Ward et al 2012, Kimber et al 2014, Minissale et al 2016). As a consequence, the diffusion of O is slower and surface reactions with atomic oxygen are less efficient. Fig. 3 shows the modelled abundance of a selection of species as a function of time in the ices (sum of the surface and the mantle) computed with the two sets of binding energies. The $O_{ice}$ abundance is increased by more than two orders of magnitude at all times. Another species that is strongly enhanced in the ices is the $CH_3$ radical. Despite the increase of both $O_{ice}$ and $CH_{3ice}$, the complex organic molecules such as $CH_3CHO$, $HCOOCH_3$ and $CH_3OCH_3$ are strongly decreased since the precursors (O, $CH_3$ and HCO) are less mobile with the new binding energies.

In the current version of the code, all species diffuse on the surfaces by thermal hopping only. To explain their experimental results between 6 and 25 K, Minissale et al. (2013) have proposed that oxygen atoms would diffuse by quantum tunneling at very low temperature. Based on this hypothesis, we have included the diffusion of atomic oxygen by quantum tunneling on the surfaces using equation 10 of Hasegawa et al. (1992) with a 1 Å rectangular barrier (the height of the barrier is $0.4E_D$ = 640 K at the surface). The species abundances computed with this model are also shown in Fig. 3. It appears that the diffusion of oxygen atoms by quantum tunneling does not impact strongly the COMs abundance in the ices. We also tested with the barrier width and height proposed by Minissale et al. (2013) (0.7 Å and 520 K respectively) but the results are not much sensitive to this change. The $CH_3OCH_3$ and $CH_3CHO$ gas-phase abundances are still strongly underproduced as compared to the observations (Bacmann et al. 2012; Cernicharo et al. 2012; Vastel et al. 2014) both due to the new binding energies (reducing their formation on the surface) and the new chemical desorption mechanism (desorbing less of these species). This underestimation is enhanced by the fact that methanol is much less abundant in the gas phase due to lower desorption efficiency leading to lower efficiency for COMs synthesis through gas phase methanol reactions (such as $CH_3OH + OH \rightarrow CH_3O + H_2O$ (Shannon et al. 2013) following by $CH_3O + CH_3 \rightarrow CH_3OCH_3 + h\nu$ (Vasyunin & Herbst 2013)).

The $CO_2$ ice abundance is only decreased by a factor of a few by the new binding energy of O but allowing the quantum tunnelling diffusion of O removes this effect. The impact of the new binding energies is limited on the methanol abundance (as on its precursors in the ices). All models we have presented here have been obtained with a fraction of the diffusion energy to the binding energy at the surface of 0.4 following Ruaud et al. (2016). Assuming a larger fraction, i.e. a less efficient diffusion, the model results are more sensitive to the new binding energies.

## 4 CONCLUSIONS

In this paper, we have studied the impact of using the new formalism proposed by Minissale et al. (2016) to quantify the efficiency of chemical desorption during reactions occurring at the surface of the grains in cold cores. A key parameter for this formalism is the binding energy of the product to the surface. We then also propose a new way to estimate binding energy for atoms, molecules and radicals. The new binding energies computed with this method are given in Table 2 in the appendix and can be used in astrochemical models.

The use of the Minissale et al. (2016) chemical desorption model leads to a much smaller gas-phase methanol abundance as compared to what is observed in cold cores and as compared to what was obtained with Garrod et al. (2007) chemical desorption model. The gas phase abundances of complex organic molecules are also much smaller with the new model (including the new chemical desorption formalism and the new binding energies), worsening the agreement between models and observations. Since methanol and other COMs formation involves grains (either directly or indirectly by the formation of precursors on grains), this underestimation of the abundances suggests either that the formation of these species on the surface is more efficient or that the chemical desorption is more efficient than suggested by Minissale et al. (2016) for some systems. We can also miss another non-thermal desorption mechanism. Clearly more work is needed to describe the desorption of molecules in cold interstellar medium.

## ACKNOWLEDGEMENTS

V.W. et M.R. thank the following funding agencies for their partial support of this work: the French CNRS/INSU programme PCMI and the ERC Starting Grant (3DICE, grant agreement 336474).

## REFERENCES

Al-Halabi A., van Dishoeck E. F., 2007, MNRAS, 382, 1648
Bacmann A., Taquet V., Faure A., Kahane C., Ceccarelli C., 2012, A&A, 541, L12
Bahr S., Toubin C., Kempter V., 2008, J. Chem. Phys., 128, 134712
Bergeat A., Moisan S., Méreau R., et al., 2009, Chem. Phys. Lett., 480, 21
Bertin M., et al., 2012, Physical Chemistry Chemical Physics (Incorporating Faraday Transactions), 14, 9929
Bertin M., et al., 2013, ApJ, 779, 120
Buch V., Czerminski R., 1991, J. Chem. Phys., 95, 6026
Burke D. J., Brown W. A., 2010, Physical Chemistry Chemical Physics (Incorporating Faraday Transactions), 12, 5947
Cernicharo J., Marcelino N., Roueff E., Gerin M., Jiménez-Escobar A., Muñoz Caro G. M., 2012, ApJ, 759, L43
Collings M. P., Dever J. W., Fraser H. J., McCoustra M. R. S., 2003, Ap&SS, 285, 633
Collings M. P., Anderson M. A., Chen R., Dever J. W., Viti S., Williams D. A., McCoustra M. R. S.,

2004, MNRAS, 354, 1133
Cuppen H. M., Herbst E., 2007, ApJ, 668, 294
Enrique-Romero J., Rimola A., Ceccarelli C., et al., 2016, MNRAS Letters, 459, L6
Fayolle E. C., Bertin M., Romanzin C., Michaut X., Öberg K. I., Linnartz H., Fillion J.-H., 2011, ApJ, 739, L36
Fayolle E. C., et al., 2013, A&A, 556, A122
Fraser H. J., Collings M. P., McCoustra M. R. S., Williams D. A., 2001, MNRAS, 327, 1165
Garrod R. T., Herbst E., 2006, A&A, 457, 927
Garrod R., Park I. H., Caselli P., Herbst E., 2006, Faraday Discussions, 133, 51
Garrod R. T., Wakelam V., Herbst E., 2007, A&A, 467, 1103
Geppert W. D., et al., 2006, Faraday Discussions, 133, 177
Hama T., Watanabe N., 2013, Chem. Rev., 113, 8783
Hasegawa T. I., Herbst E., 1993, MNRAS, 261, 83
Hasegawa T. I., Herbst E., Leung C. M., 1992, ApJS, 82, 167
He J., Shi J., Hopkins T., Vidali G., Kaufman Michael J., 2015, ApJ, 801, 120
Hickson K.M., Caubet P., Loison J.-C., 2013, J. Phys. Chem. Lett., 4, 2843
Kassel L. S., 1928, J. Phys. Chem., 32, 225
Kimber H.J., Ennis C.P., Price S.D., 2014, Faraday Discussions, 168, 167
Lattelais M., et al., 2011, A&A, 532, A12
Leger A., Jura M., Omont A., 1985, A&A, 144, 147
Luca A., Voulot D., Gerlich D., 2002, Proceedings of Contributed Papers, PART II, 204
Minissale M., et al., 2013, Physical Review Letters, 111, 053201
Minissale M., Dulieu F., Cazaux S., Hocuk S., 2016, A&A, 585, A24
Muñoz Caro G. M., Jiménez-Escobar A., Martín-Gago J. Á., Rogero C., Atienza C., Puertas S., J.M. Sobrado, J. Torres-Redondo 2010, A&A, 522, p. A108
Noble J. A., Theule P., Mispelaer F., Duvernay F., Danger G., Congiu E., Dulieu F., Chiavassa T., 2012, A&A, 543, A5
Noble J. A., et al., 2015, A&A, 576, A91
Olanrewaju B. O., Herring-Captain J., Grieves G. A., Aleksandrov A., Orlando T. M., 2011, The Journal of Physical Chemistry A, 115, 5936
Perets H. B., Biham O., Manico G., Pirronello V., Roser J., Swords S., Vidali G., 2005, ApJ, 627, 850
Pratap P., Dickens J. E., Snell R. L., Miralles M. P., Bergin E. A., Irvine W. M., Schloerb F. P., 1997, ApJ, 486, 862
Raut U., Famà M., Teolis B.D., et al., 2007, J. Chem. Phys., 127, 204713
Rice O. K., Ramsperger H. C., 1927, J. Am. Chem. Soc., 49, 1617
Roberts J. F., Rawlings J. M. C., Viti S., Williams D. A., 2007, MNRAS, 382, 733
Ruaud M., Wakelam V., Hersant F., 2016, MNRAS, 459, 3756
Sandford S. A., Allamandola L. J., 1990, Icarus, 87, 188
Sandford S. A., Allamandola L. J., 1993, ApJ, 409, L65
Shen C. J., Greenberg J. M., Schutte W. A., van Dishoeck E. F., 2004, A&A, 415, 203
Shannon R.J., Blitz M.A., Goddard A., et al., 2013, Nature Chemistry, 5, 745
Szori M., Jedlovszky P., 2014, The Journal of Physical Chemistry C, 118, 3599
Tielens A. G. G. M., Hagen W., 1982, A&A, 114, 245
Vastel C., Ceccarelli C., Lefloch B., Bachiller R., 2014, ApJ, 795, L2
Vasyunin A. I., Herbst E., 2013, ApJ, 769, 34
Wakelam V., Vastel C., Aikawa Y., Coutens A., Bottinelli S., Caux E., 2014, MNRAS, 445, 2854
Wakelam V., et al., 2015, ApJS, 217, 20
Ward M. D., Hogg I. A., Price S. D., 2012, MNRAS, 425, 1264
Wang J.-H., Han K.-L., He G.-Z., et al., 2003, J. Phys. Chem. A, 107, 9825
Ward M.D., Hogg I.A., Price S.D., 2012, MNRAS, 425, 1264
Willacy K., Millar T. J., 1998, MNRAS, 298, 562
Zhao Y., Truhlar D., 2008, Theor. Chem. Acc., 120, 215
Zubkov T., Smith R. S., Engstrom T. R., Kay B. D., 2007, The Journal of Chemical Physics, 127,

184707

**Table 1:** Percentage of products chemically desorbed during surface reactions computed with the two formalisms. The values for RRK have been computed with $a = 0.01$.

| Reaction | RRK | MDCH |
|---|---|---|
| $O + H \rightarrow OH$ | 0.9 | 30 |
| $O + OH \rightarrow H_2O$ | 0.7 | 0.25 |
| $N + H \rightarrow NH$ | 0.9 | 5.5 |
| $NH + H \rightarrow NH_2$ | 0.7 | 2.6 |
| $NH_2 + H \rightarrow NH_3$ | 0.5 | 1.1 |
| $HCO + H \rightarrow H_2CO$ | 0.7 | 2 |
| $CH_3O + H \rightarrow CH_3OH$ | 0.25 | $3 \times 10^{-2}$ |
| $O + CO \rightarrow CO_2$ | 0.9 | 1.7 |
| $HS + H \rightarrow H_2S$ | 0.8 | 1.7 |
| $C_3H_3 + H \rightarrow CH_3CCH$ | 0.2 | $2 \times 10^{-3}$ |

**Table 2: Binding energies of species on amorphous water ice surface**
Column 1: Name of the species
Column 2: Drawing of the minimum energy geometry computed by our model. In some cases, there several possible minima.
Column 3: Adsorption energy computed by our model using DFT calculations corresponding to the geometry showed in the second column.
Column 4: Adsorption energy computed by our model using MP2 calculations corresponding to the geometry showed in the second column
Column 5: The energy we suggest and that we use in our chemical model. In the absence of experimental data, we use the value obtained by our model. Other wise, the proposed value is an educated guess from the different available data (including ours).
Column 6: Binding energies previously used in the model.
Column 7: Experimental (or theoretical) data published in the literature and comments in some cases.

| Specie | drawing | Model M06 (K) | Model MP2 (K) | Values used in the simulation (K) | Previous values (K) | Literature values (K) and comments |
|---|---|---|---|---|---|---|
| H | 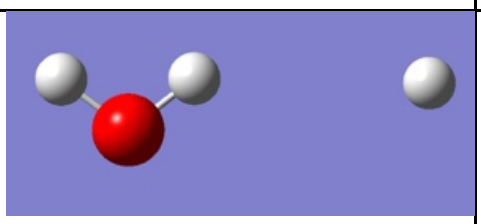 | 400 | 680 | 650 | 650 | 650 (Minissale *et al.* 2016)<br>210-430 (Buch & Czerminski 1991)<br>650±10 (Al-Halabi & Van Dishoeck 2007) |
| $H_2$ | 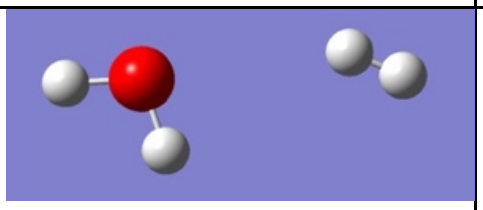 | 800 | | 440 | 440 | 555±35 (Sandford & Allamandola 1993)<br>540-840 (Perets *et al.* 2005)<br>440 (Cuppen & Herbst 2007) |
| O | 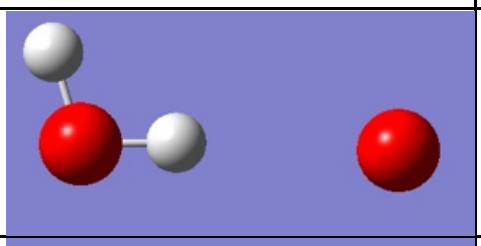 | 1700 | 1800 | 1600 | 800 | 1660±60 (He *et al.* 2015)<br>1400 (Minissale et al. 2016)<br>1680±240 (Kimber *et al.* 2014)<br>1504±12 (Ward *et al.* 2012) |
| | 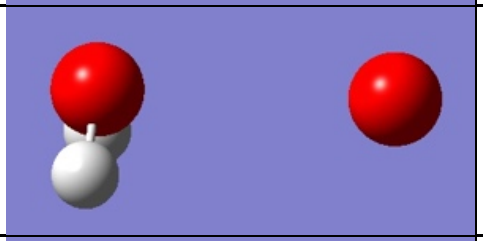 | 2200 | 1500 | | | |
| $O_2$ | 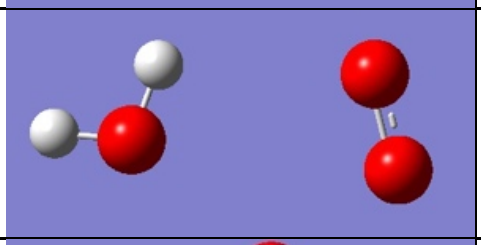 | 1000 | 900 | 1200 | 1000 | 1200 (Minissale et al. 2016)<br>1000 (Hama & Watanabe 2013) |
| $O_3$ | 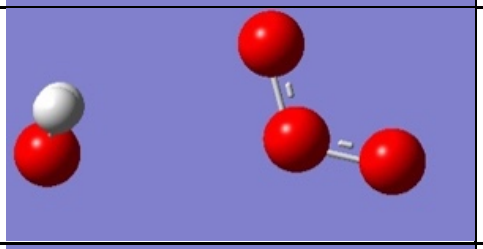 | 1500 | 1900 | 2100 | 1800 | 2100 (Minissale et al. 2016) |
| $HO_2$ | 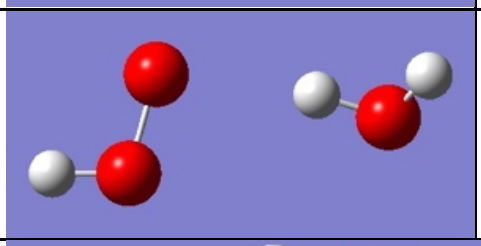 | 3000 | | 5000 | 3650 | 4000 (Minissale et al. 2016) |
| | 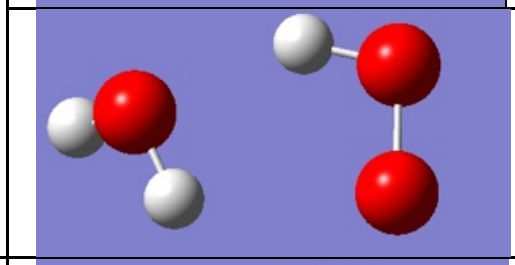 | 8500 | | | | |
| $H_2O_2$ | 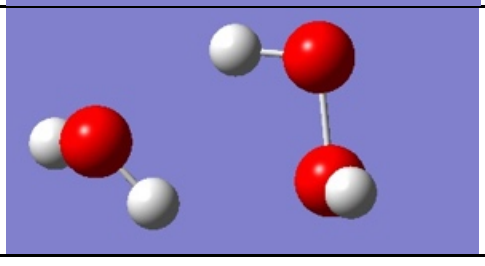 | 6800 | 6100 | 6000 | 5700 | 6000 (Minissale et al. 2016) |
| OH | 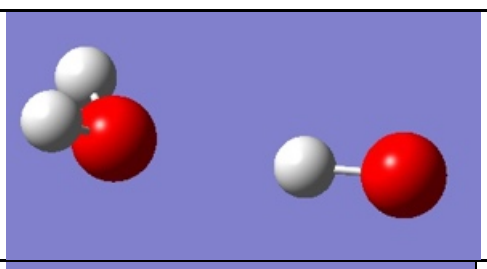 | 5300 | 4800 | 4600 | 2850 | 4600 (Minissale et al. 2016) |
| | 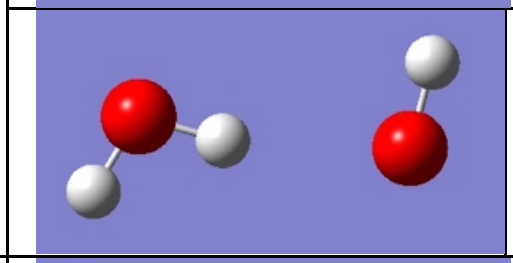 | 3300 | | | | |
| $H_2O$ | 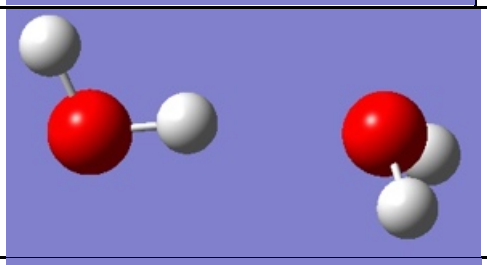 | 4600 | 3900 | 5600 | 5700 | 5600 (Fraser *et al.* 2001)<br>4815 (Sandford & Allamandola 1990) |
| C$\ell$ | 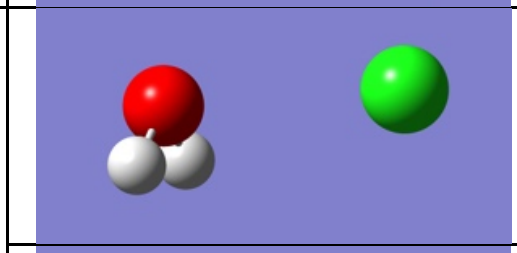 | 4400 | 3100 | 3000 | 1100 | |
| | 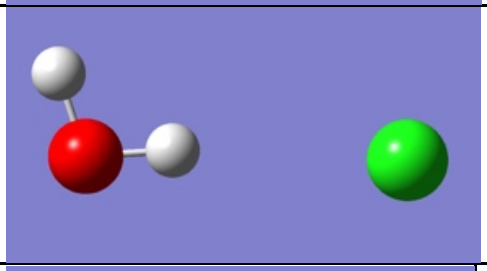 | 1300 | 1700 | | | |
| HC$\ell$ | 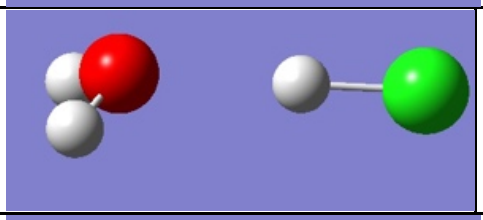 | 4800 | 5000 | 5172 | 5174 | 5172 (Olanrewaju *et al.* 2011) |
| HF | 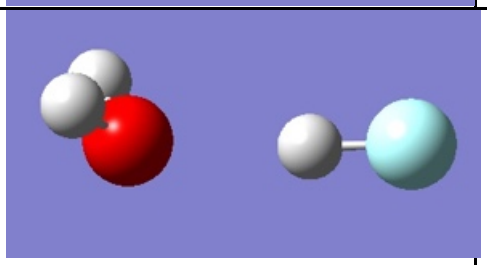 | 7900 | 7300 | 7500 | 2850 | |
| CO | 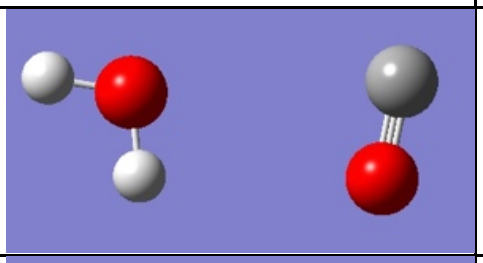 | 1300 | 1100 | 1300 | 1150 | 1300 (Minissale et al. 2016)<br>1740 (Sandford & Allamandola 1990)<br>1180 (Collings *et al.* 2003) |
| HCO | 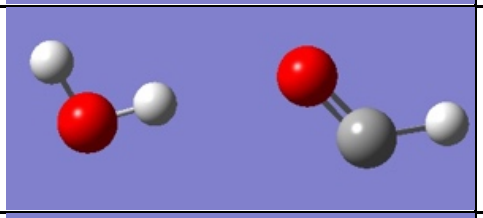 | 2300 | 2100 | 2400 | 1600 | 1600 (Minissale et al. 2016)<br>2333 (Enrique-Romero *et al.* 2016) (water cluster calculations) |
| | 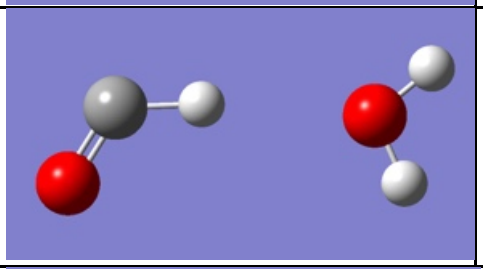 | 2700 | 2600 | | | |
| $H_2CO$ | 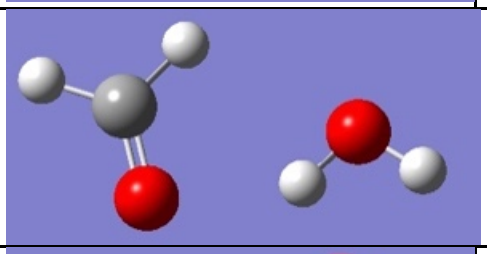 | 5100 | | 4500 | 2050 | 3200 (Noble *et al.* 2012) |
| $CH_3O$ | 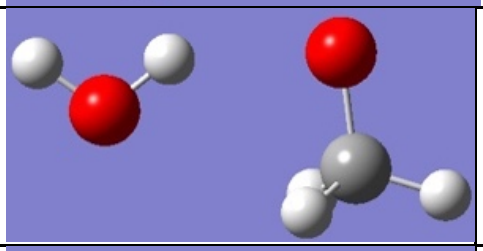 | 4700 | | 4400 | 5084 | 3700 (Minissale et al. 2016) |
| $CH_2OH$ | 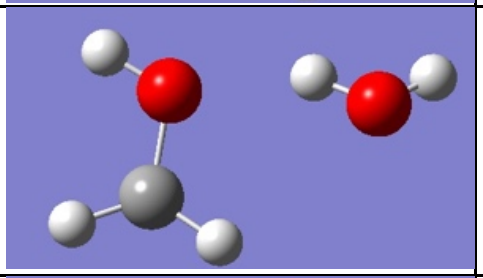 | 3900 | | 4400 | 5084 | |
| | 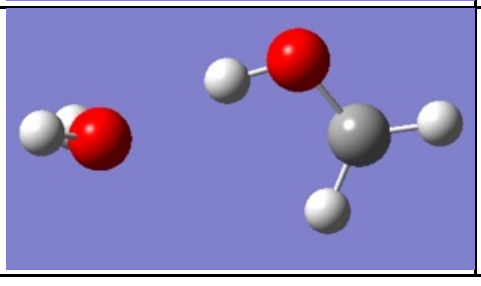 | 5800 | | | | |
| $CH_3OH$ | 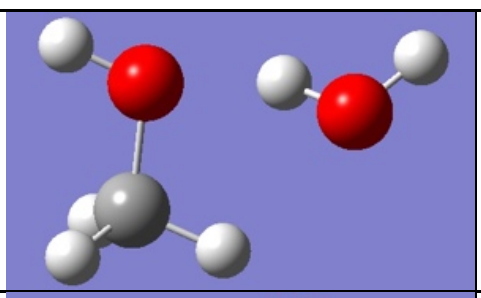 | 5100 | 5100 | 5000 | 5534 | 3700 (Minissale et al. 2016)<br>5400 (Bahr *et al.* 2008)<br>5534 (Collings *et al.* 2004) |
| | 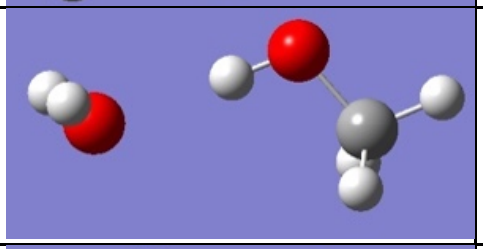 | 4500 | 4600 | | | |
| $C_2H_5OH$ | 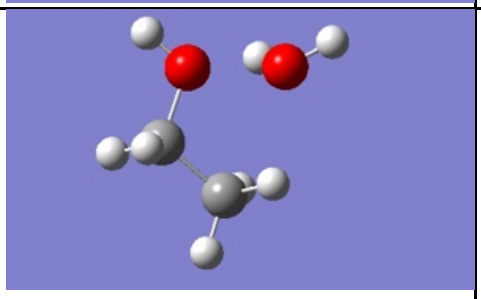 | 5500 | | 5400 | 6584 | 6800 (theory, crystalline Ice) (Lattelais *et al.* 2011) |
| | 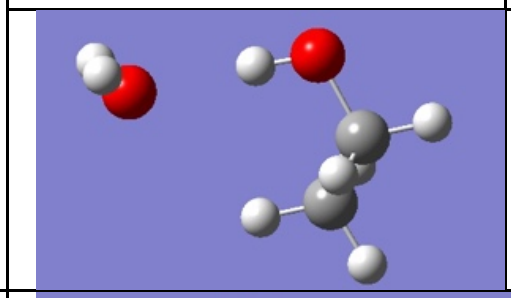 | 4200 | | | | |
| $CH_3CHO$ | 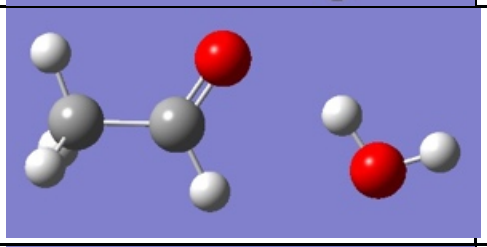 | 5400 | | 5400 | 2450 | |
| $C_2H_3CHO$ | 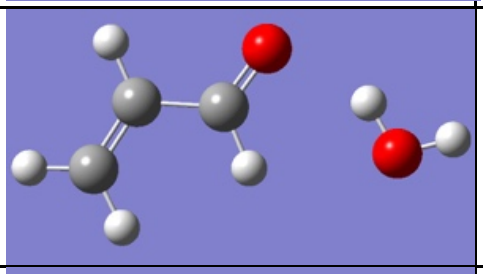 | 5400 | | 5400 | | |
| $C_2H_5CHO$ | 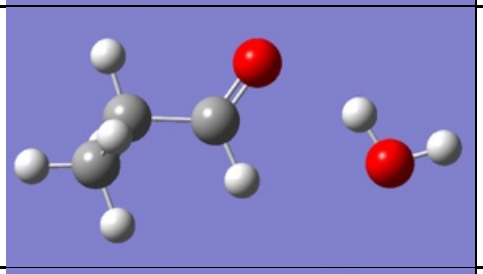 | 4500 | | 4500 | | |
| c-$C_3H_6O$ | 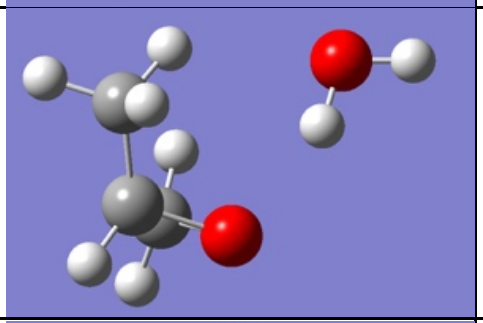 | 5600 | | 5600 | | |
| $H_2CCO$ | 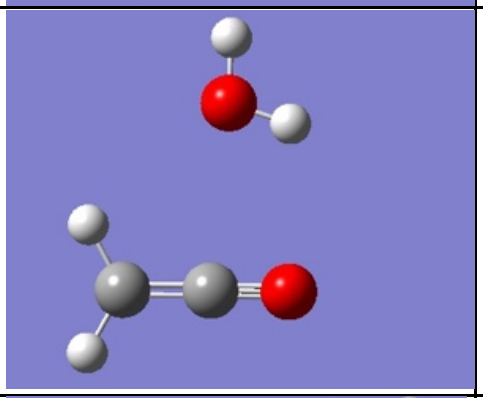 | 3000 | | 2800 | 2200 | |
| | 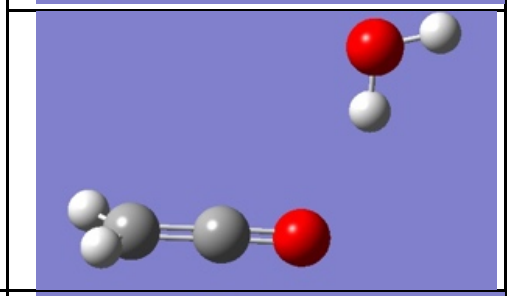 | 2300 | | | | |
| $CO_2$ | 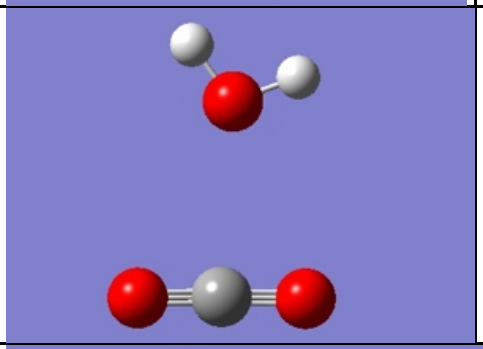 | 3100 | 2600 | 2600 | 2575 | 2300 (Minissale et al. 2016)<br>2400 (Hama & Watanabe 2013)<br>2860 (Sandford & Allamandola 1990)<br>2575 (Collings et al. 2004) |
| OCS | 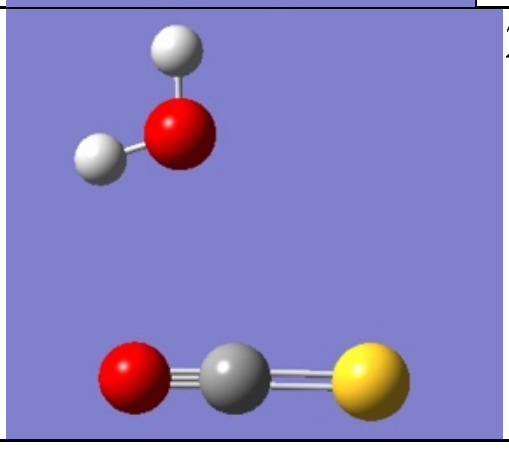 | 2100 | | 2400 | 2888 | 2430 (Ward et al. 2012)<br>2888 (Collings et al. 2004) |
| N | 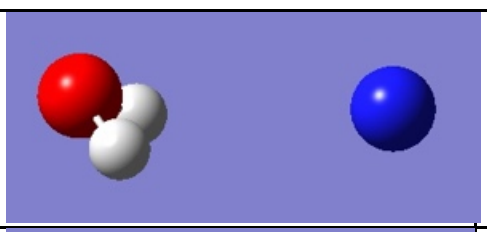 | 1200 | 900 | 720 | 800 | 720 (Minissale et al. 2016) |
| NH | 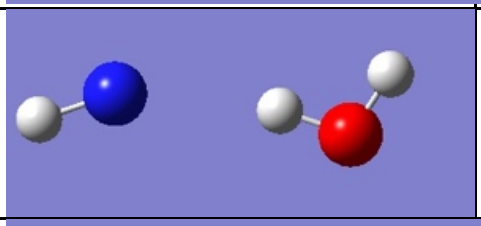 | 2800 | 2400 | 2600 | 2378 | |
| | 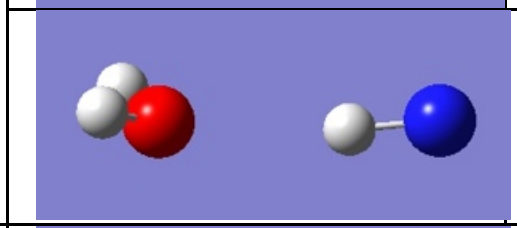 | 3000 | 2800 | | | |
| $NH_2$ | 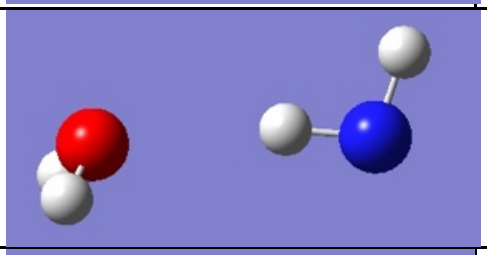 | 2800 | 2500 | 3200 | 3956 | |
| | 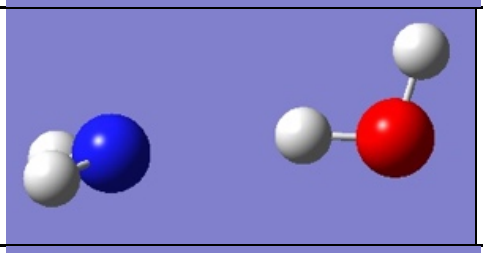 | 4500 | 4100 | | | |
| $NH_3$ | 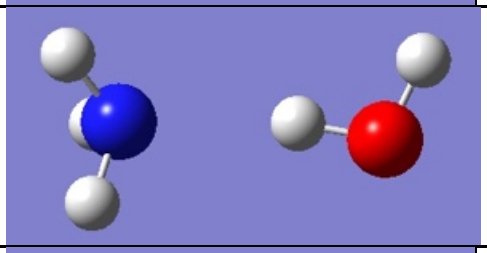 | 5600 | 5500 | 5500 | 5534 | 5500 in (Hama & Watanabe 2013) 5534 (Collings et al. 2004) |
| CN | 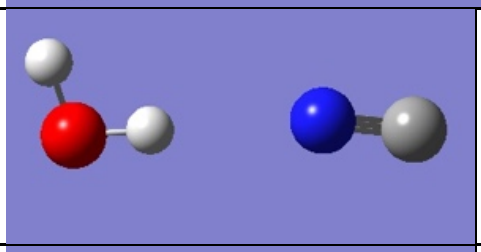 | 2500 | 3000 | 2800 | 1600 | |
| HCN | 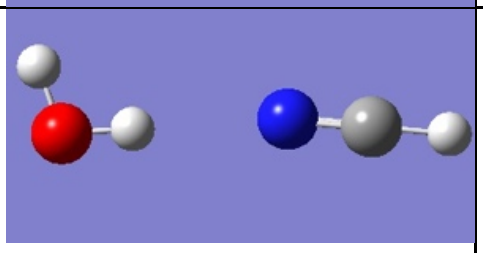 | 3500 | 3900 | 3700 | 2050 | It should be noted that Monte Carlo simulation study leads to a much larger value, around 4700 K (Szőri & Jedlovszky 2014) |
| HNC | 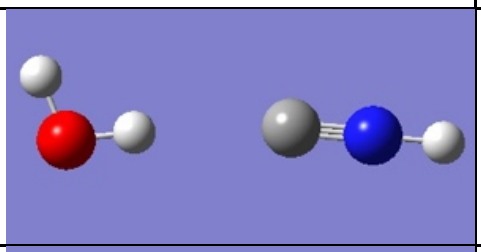 | 3600 | 4000 | 3800 | 2050 | |
| $CH_3CN$ | 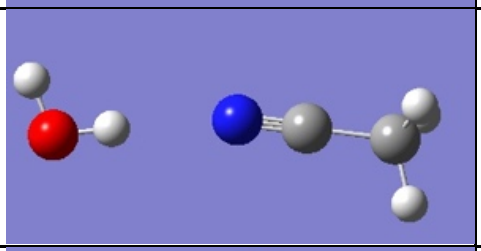 | 4300 | 4800 | 4680 | 4680 | 4680 (Collings et al. 2004) |
| $NH_2CHO$ | 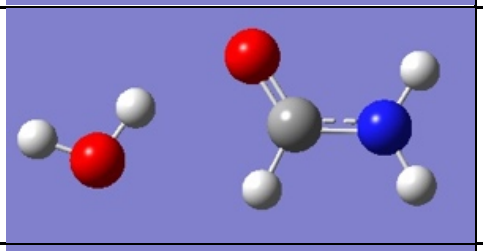 | 6300 | | 6300 | 5556 | |
| HNCO | 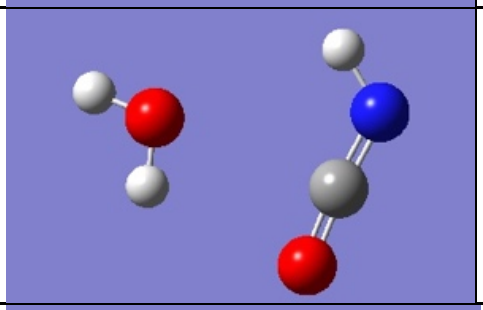 | 4000 | | 4400 | 2850 | 3900 (Noble *et al.* 2015) |
| | 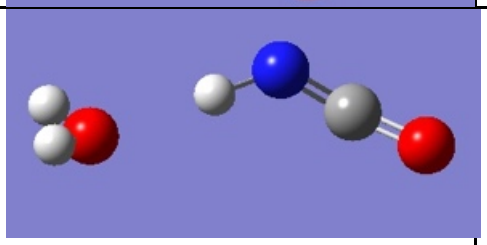 | 5700 | | | | |
| HNCS | 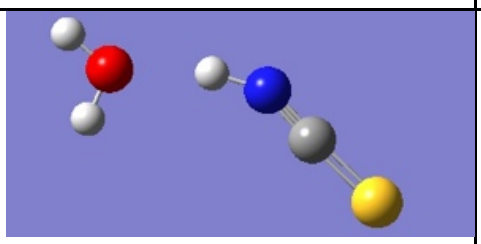 | 5800 | | 4600 | - | We use a value similar to HNCO considering the experimental and the Fsem result for HNCO and HNCS |
| $CH_3NCO$ | 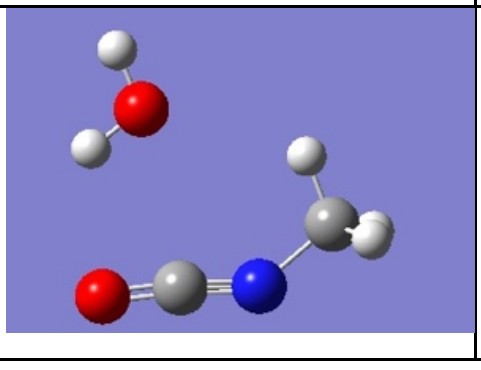 | 4700 | | 4700 | - | |
| $CH_2NCHO$ | 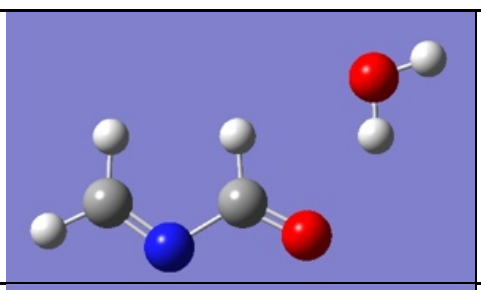 | 5300 | | 5300 | - | |
| NO | 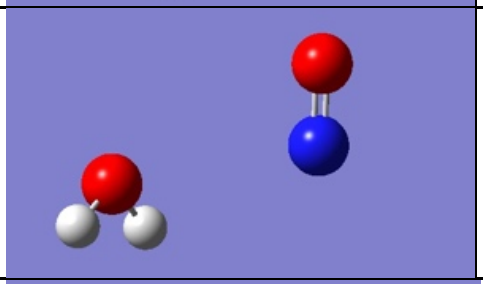 | 1600 | 1700 | 1600 | 1600 | |
| $N_2$ | 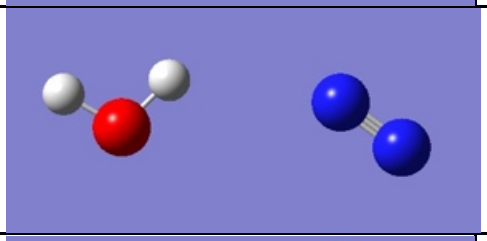 | 1100 | 1400 | 1100 | 1000 | 1000-1600 (Zubkov *et al.* 2007a, Zubkov *et al.* 2007b)<br>810-1150 in (Minissale et al. 2016) |
| HNO | 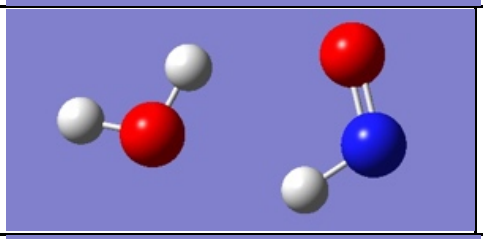 | 5000 | | 3000 | 2050 | H bonding likely overestimated when HNO acts as a donor and an acceptor considering ASW structure. |
| | 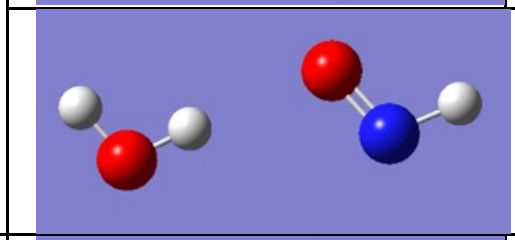 | 2500 | | | | |
| $NH_2OH$ | 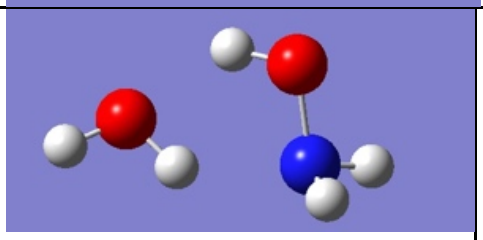 | 7400 | | 6000 | - | H bonding likely overestimated as $NH_2OH$ acts as a donor and an acceptor, considering ASW structure |
| C | 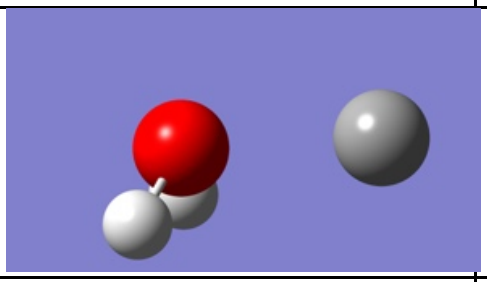 | 10000 | | 10000 | 4000 | |
| CH | No complex | | | | 925 | Direct insertion (Bergeat *et al.* 2009, Hickson *et al.* 2013) |
| $CH_2$ | 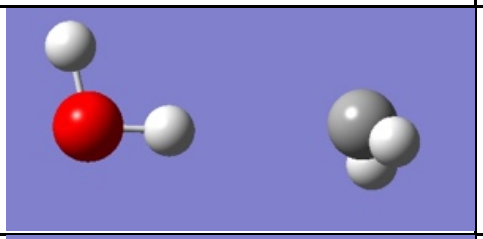 | 2600 | 1300 | 1400 | 1050 | |
| $CH_3$ | 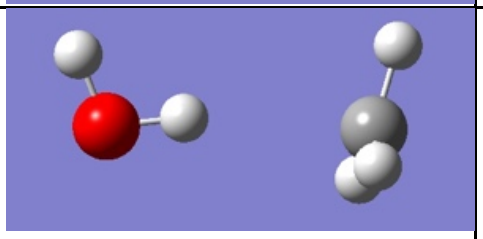 | 2500 | 1500 | 1600 | 1175 | 734 (Enrique-Romero et al. 2016) (water cluster calculations) |
| $CH_4$ | 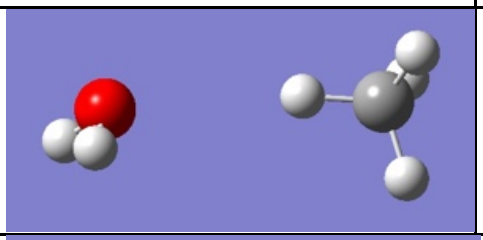 | 800 | 1000 | 960 | 1300 | 960 (Raut *et al.* 2007) |
| $C_2$ | 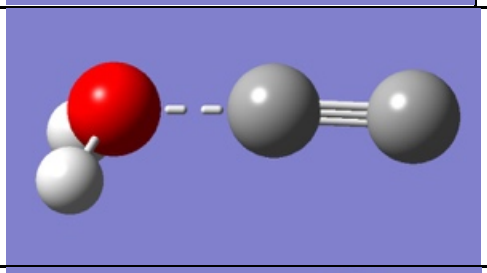 | 24000 | | 10000 | 1600 | The stability of $CCOH_2$ species is highly dependent of the method used for the calculation, see also (Wang *et al.* 2003). |
| $C_2H$ | 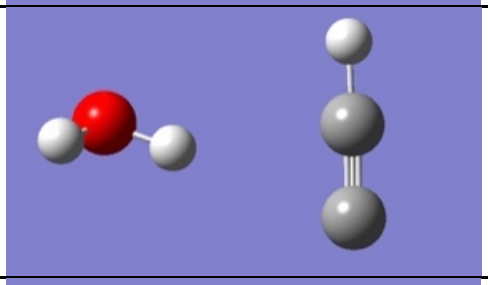 | 2400 | | 3000 | 2137 | |
| | 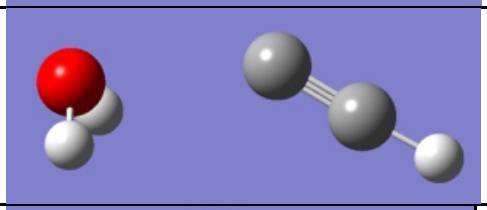 | 3400 | | | | |
| $C_2H_2$ | 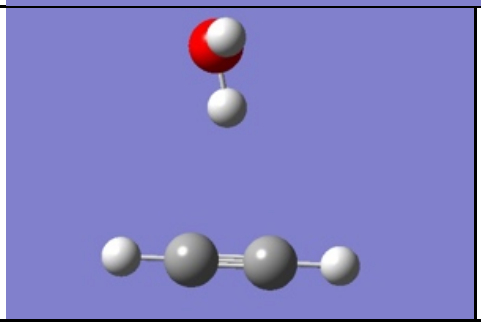 | 2600 | 2700 | 2587 | 2587 | 2587 (Collings et al. 2004) |
| $C_2H_3$ | 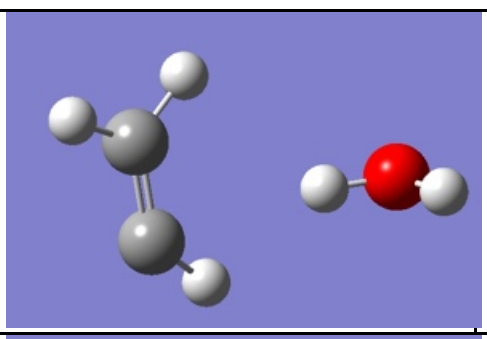 | 2600 | | 2800 | 3037 | |
| | 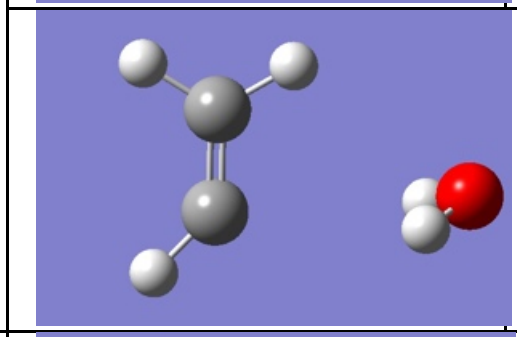 | 3000 | | | | |
| $C_2H_4$ | 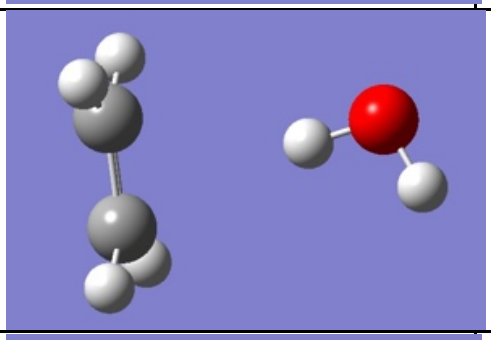 | 2500 | | 2500 | 3487 | |
| $C_2H_5$ | 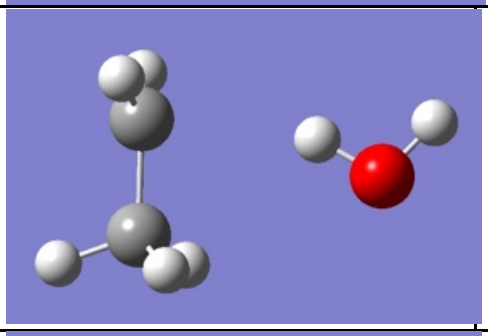 | 3100 | | 3100 | 3937 | |
| $C_2H_6$ | 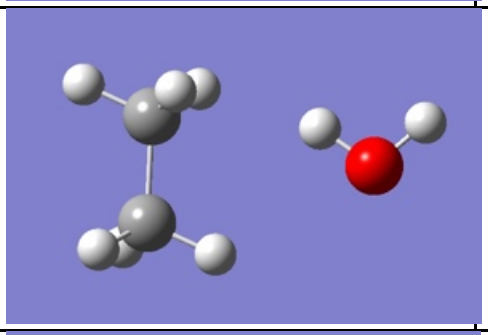 | 1600 | | 1600 | 4387 | |
| $C_3$ | 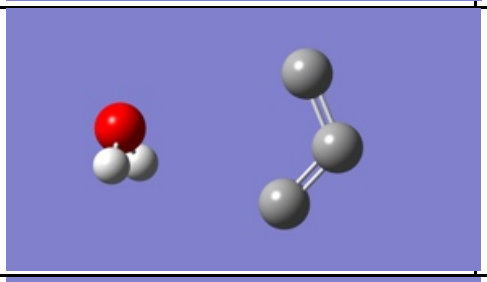 | 2500 | | 2500 | 2400 | |
| l-$C_3H$ | 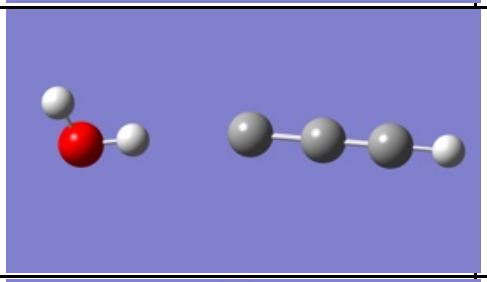 | 4000 | | 4000 | 2937 | |
| c-$C_3H$ | 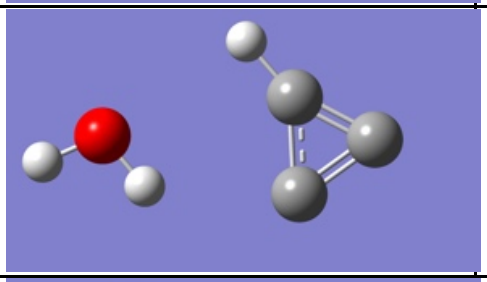 | 5200 | | 5200 | 2937 | |
| c-$C_3H_2$ | 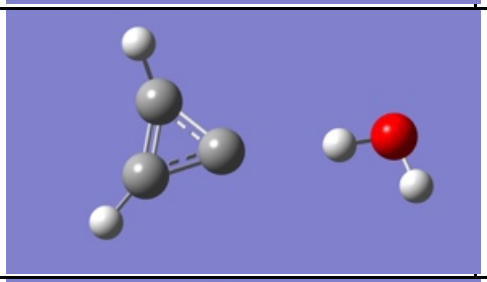 | 5900 | | 5900 | 3387 | |
| $C_3H_3$ | 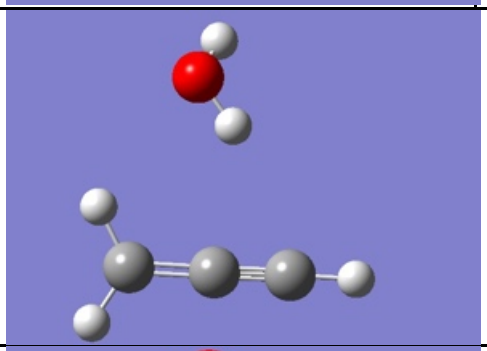 | 3300 | | 3300 | 3837 | |
| $CH_3CCH$ | 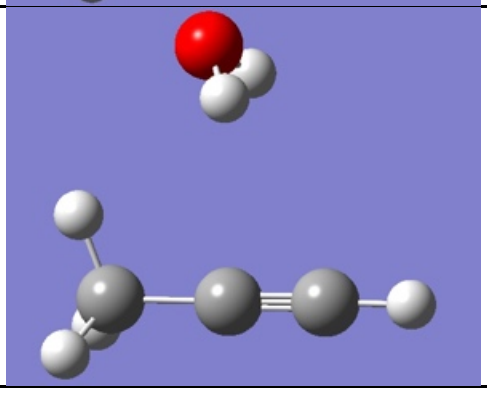 | 3800 | 3800 | 3800 | 4287 | 2500±40 (Kimber et al. 2014). |
| $H_2CCCH_2$ | 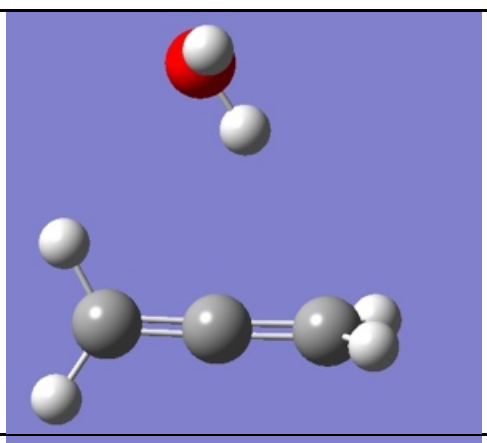 | 3000 | | 3000 | - | |
| $C_3H_5$ | 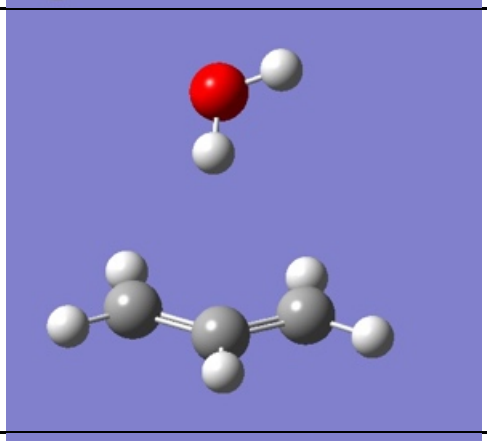 | 3100 | | 3100 | - | |
| $C_3H_6$ | 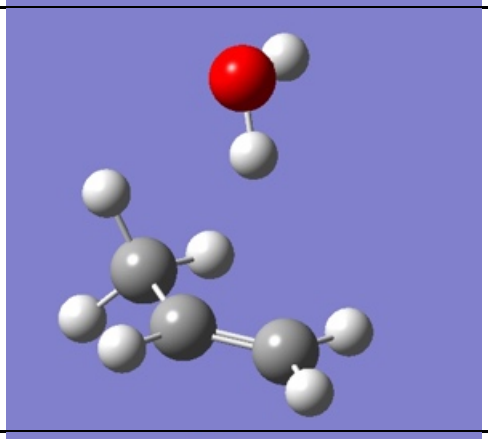 | 3100 | | 3100 | - | |
| $C_3H_7$ | 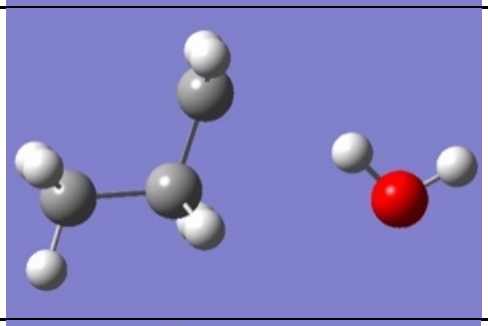 | 3100 | | 3100 | - | |
| $C_3H_8$ | 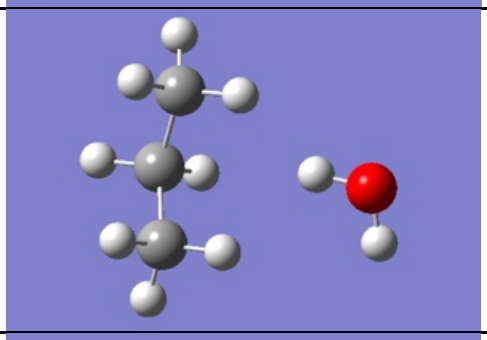 | 1600 | | 1600 | - | |
| S | 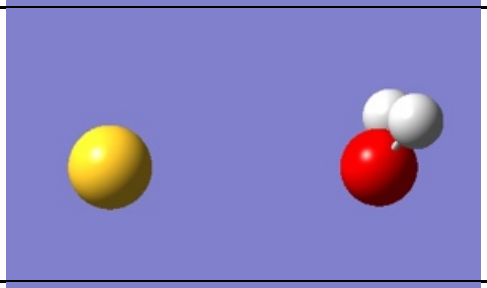 | 2600 | | 2600 | 1100 | |
| HS | 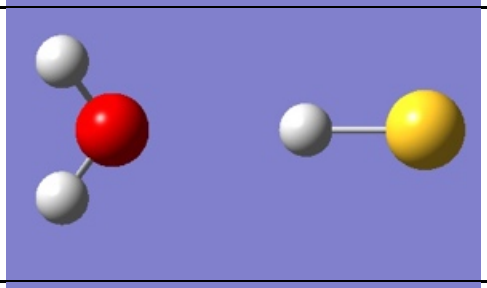 | 2700 | | 2700 | 1450 | |
| | 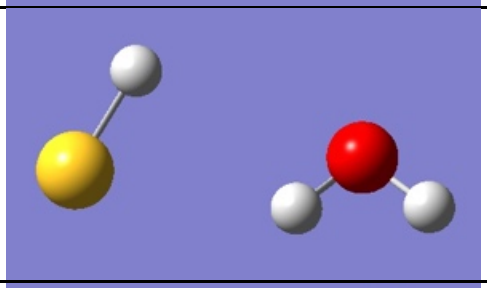 | 2700 | | | | |
| $H_2S$ | 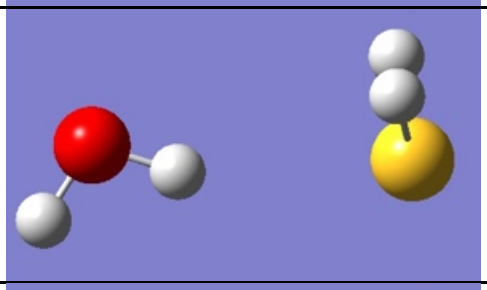 | 2900 | 2700 | 2700 | 2743 | 2743 (Collings et al. 2004) |
| | 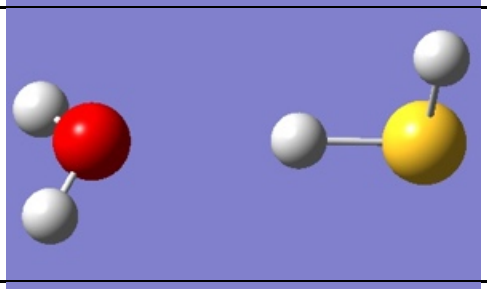 | 2500 | 2400 | | | |
| CS | 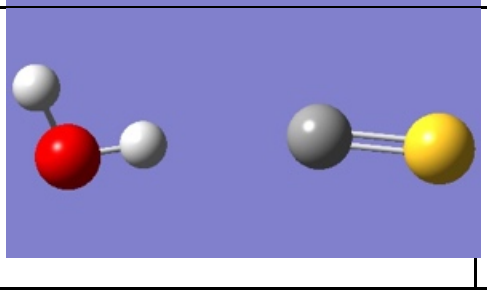 | 3200 | | 3200 | 1900 | |
| HCS | 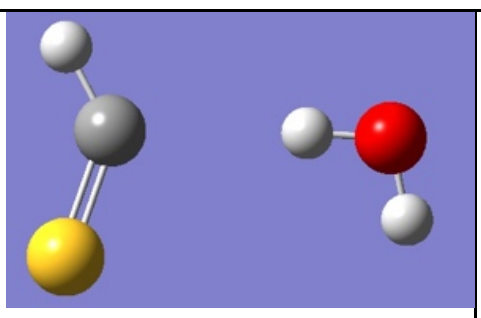 | 2900 | | 2900 | 2350 | |
| $H_2CS$ | 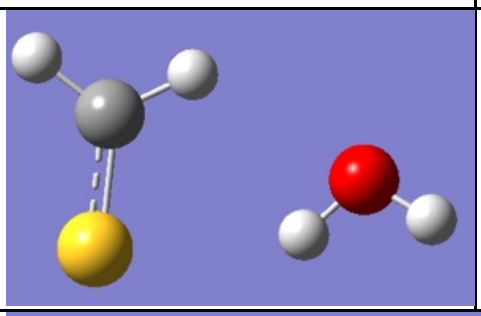 | 4400 | | 4400 | 2700 | |
| $CH_3S$ | 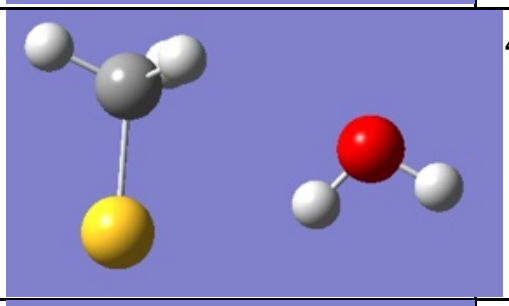 | 4200 | | 4200 | - | |
| $CH_2SH$ | 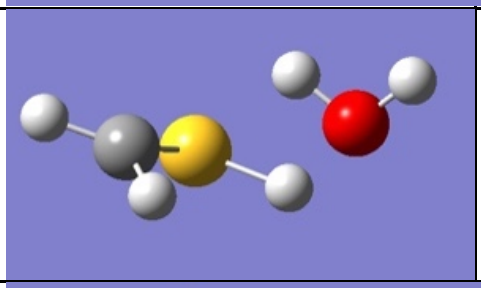 | 3700 | | 3700 | - | |
| $CH_3SH$ | 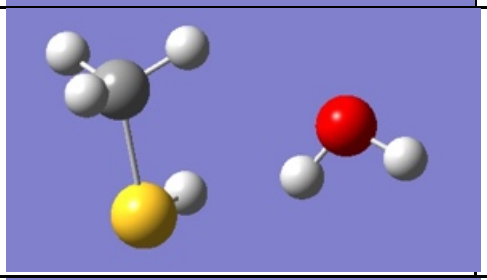 | 4200 | | 4000 | - | |
| | 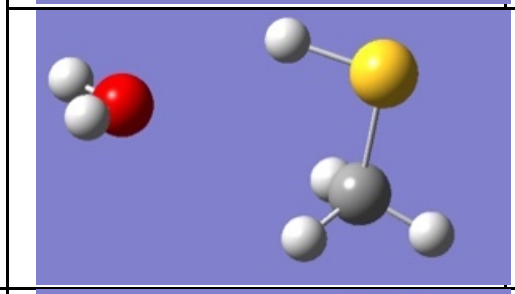 | 2300 | | | | |
| SO | 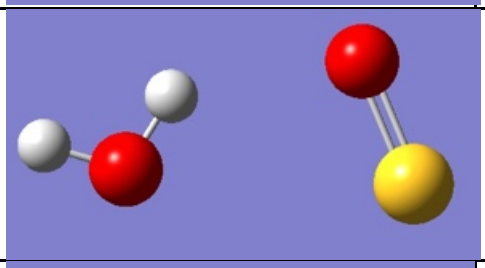 | 2900 | | 2800 | 2600 | |
| | 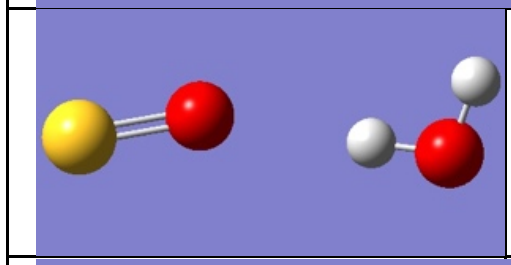 | 2400 | | | | |
| | 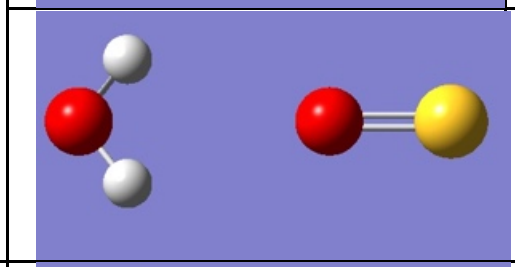 | 2000 | | | | |
| $SO_2$ | 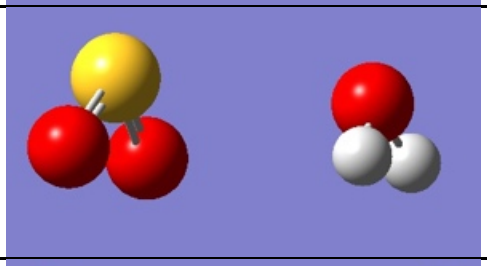 | 5000 | | 3400 | 3405 | 3405 (Collings et al. 2004) |
| | 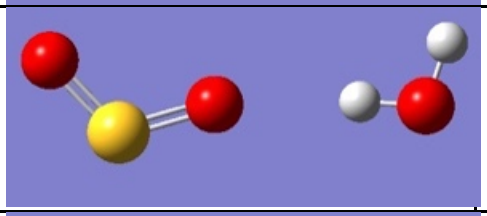 | 2400 | | | | |
| Si | 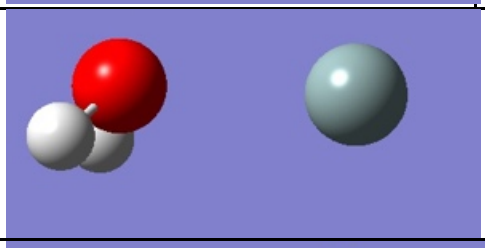 | 11600 | | 11600 | 2700 | Very strong interaction as for C…$H_2O$. |
| SiH | 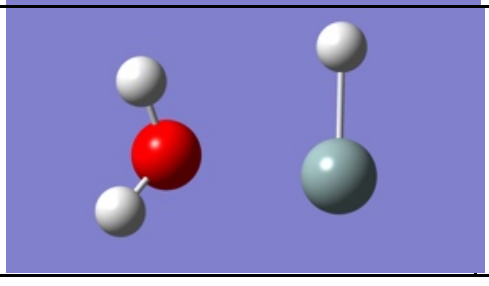 | 13000 | | 13000 | 3150 | Very strong interaction. There is a real barrier toward insertion contrary to CH + $H_2O$ reaction. |

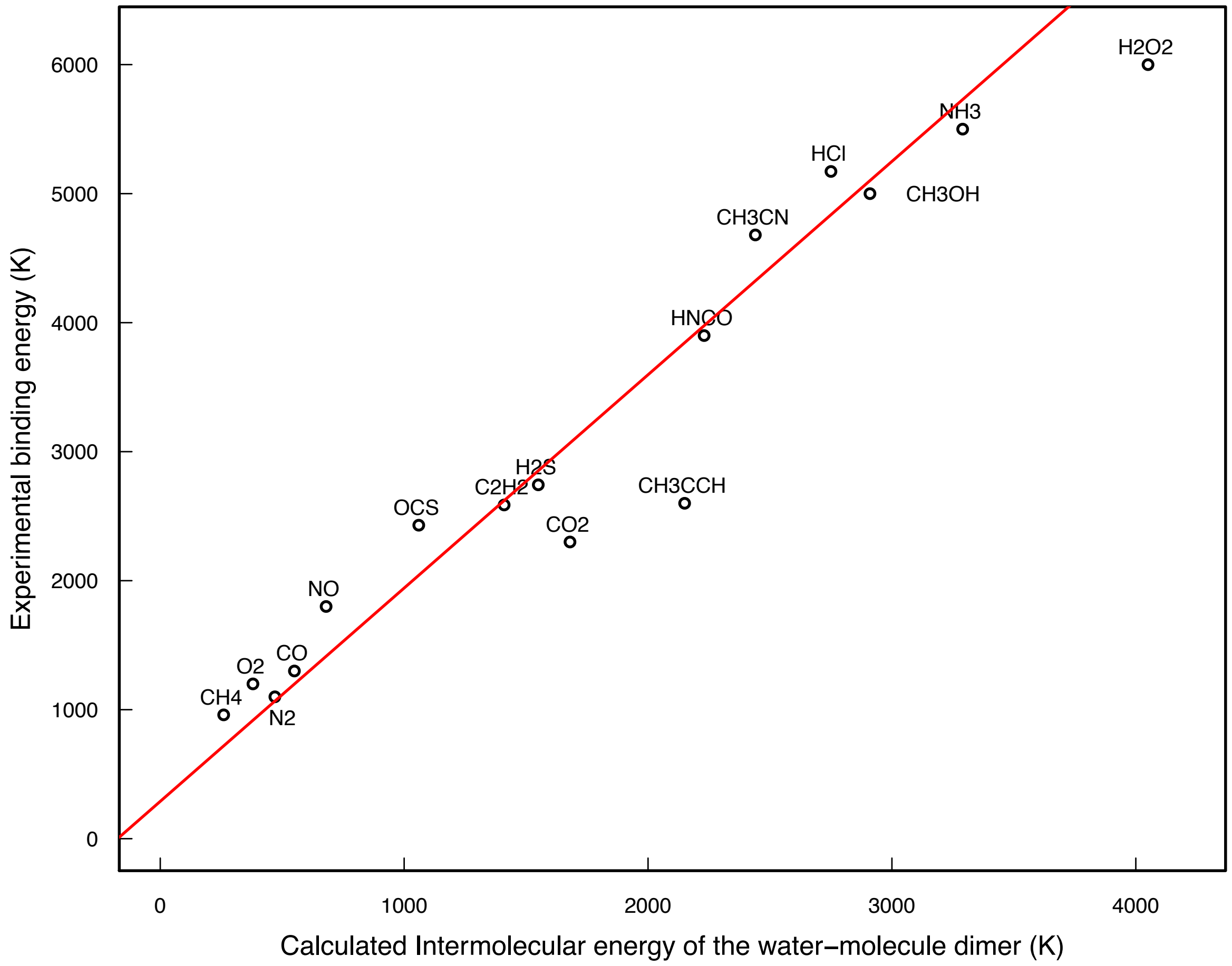

**Fig. 1** Experimental binding energies (see Table 2 for the references) versus the calculated intermolecular energies of the complexes of the various species (name on the graph) with one water molecule.

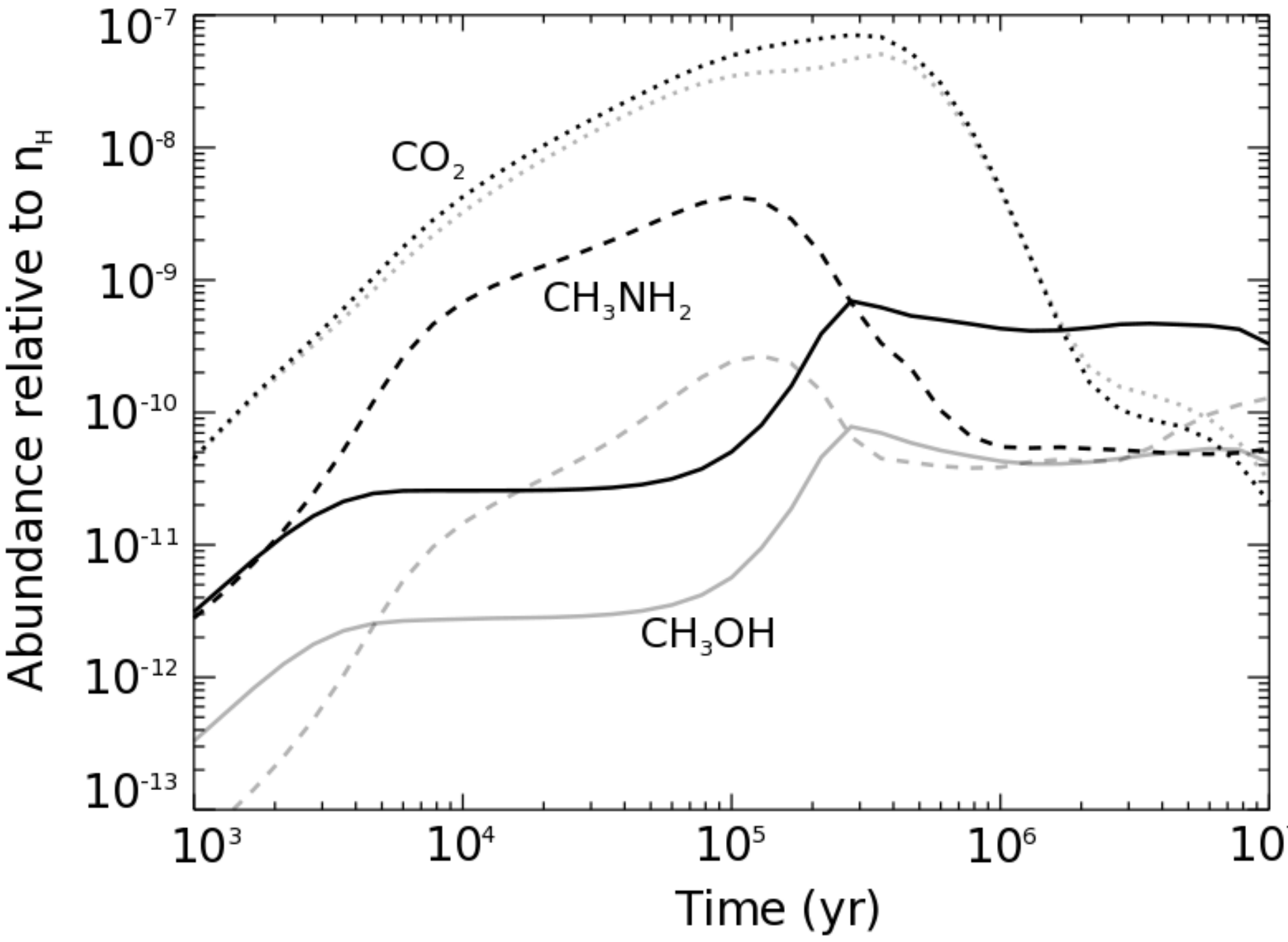

**Fig. 2** Gas phase abundances of $CH_3OH$ (solid lines), $CH_3CNH_2$ (dashed lines), and $CO_2$ (dotted lines) computed with the RRK (black lines) or the MDCH (grey lines) formalisms as a function of time.

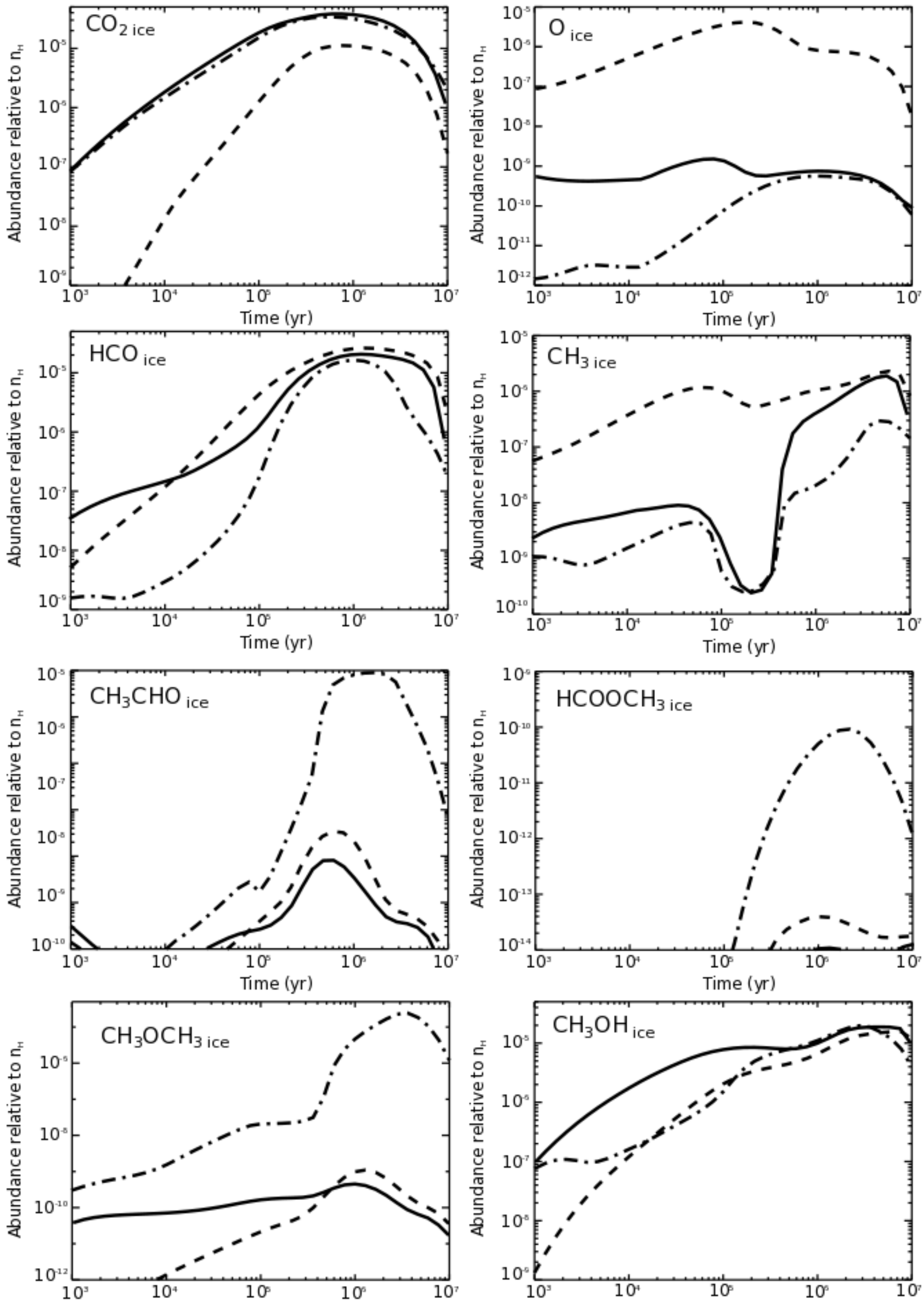

**Fig. 3** Ice abundance (sum of surface and mantle) of a selection of molecules as a function of time. In each panel, the three lines represent the abundance computed with the old binding energies (dashed-dotted) and the new binding energies with (solid) and without (dashed) the diffusion of O by quantum tunnelling.